\documentclass[runningheads]{svmult}

\usepackage{makeidx}   
\usepackage{graphicx}  
\usepackage{subeqnar}  
\usepackage{multicol}  
\usepackage{physprbb}  
\makeindex             

\begin{document}
\title*{Problems and Progress in Astrophysical Dynamos}
%
%
%
%
\titlerunning{Astrophysical Dynamos}
%
\author{Ethan T. Vishniac\inst{1}
\and A. Lazarian\inst{2}
\and Jungyeon Cho\inst{2}}

\authorrunning{Vishniac, Lazarian \& Cho}
%
%
\institute{Johns Hopkins Univ., Baltimore MD21218, USA
\and Univ. of Wisconsin, Madison WI53706, USA}

\maketitle              

\begin{abstract}
Astrophysical objects with negligible resistivity are often threaded by 
large scale magnetic fields.  The generation of these fields is 
somewhat mysterious, since a magnetic field in a perfectly conducting 
fluid cannot change the flux threading a fluid element, or the field topology.  
Classical dynamo theory evades this limit by assuming 
that magnetic reconnection is fast, even for vanishing resistivity, 
and that the large scale field can
be generated by the action of kinetic helicity.  Both these claims
have been severely criticized, and the latter appears to conflict
with strong theoretical arguments based on magnetic helicity
conservation and a series of numerical simulations.  Here we discuss
recent efforts to explain fast magnetic reconnection through the
topological effects of a weak stochastic magnetic field component.  We also show
how mean-field dynamo theory can be recast in a form which respects
magnetic helicity conservation, and how this changes our understanding
of astrophysical dynamos.  Finally, we comment briefly on why an
asymmetry between small scale magnetic and velocity fields is necessary
for dynamo action, and how it can arise naturally.
\end{abstract}

\section{Introduction}

Magnetic fields have played a curious role in astrophysics,
being both commonplace and poorly understood.
They are ubiquitous in ionized systems, from
the interiors of stars to the hot interstellar medium.
The magnetic energy density is typically roughly
comparable to the turbulent kinetic energy density.
In stellar interiors, this means that magnetic fields tend to
play a small role. In the interstellar medium, and in stellar coronae, 
their role is large, and consequently a matter of intense debate.  In
accretion disks the typical magnetic field energy density is
probably an order of magnitude below the ambient gas pressure
(e.g. \cite{hgb95,hgb96,shgb96,bnst95,bnst96}) 
but they play a critical role in the
outward transfer
of angular momentum and the dissipation of orbital energy.  Moreover,
in optically thin environments the presence of a strong magnetic
field can have a dramatic effect on the luminosity and spectrum
of an object.  A clear understanding of the generation and dynamics
of magnetic fields is important to astrophysics in many ways.
Unfortunately, their dynamics has not been well understood, at
least judging by the diversity of opinions found in the
literature \cite{cv91,p92,ka92,vpr93,bz95}. 
Consequently, arguments which cite
magnetic fields as a dynamically important element in any
particular object have tended to rely on phenomenology, rather
than any sort of fundamental explanation.

Fortunately, over the last ten years, and especially quite recently,
there has been significant progress in this area.  First, although
direct observations of high conductivity magnetic field dynamics are still 
restricted to the solar wind and the Sun, improvements in resolution have 
made it possible to watch magnetic fields evolve in real time
\cite{khm98}, and 
to measure the power spectrum of magnetohydrodynamic (MHD) turbulence in the 
solar
wind directly \cite{lsnm98}.
Second, numerical simulations have reached the point where it
is possible to simulate simple MHD systems with 
$\sim 10^8$ cells over many dynamical times.  Third, a
better understanding has been reached in terms of MHD turbulence theory
(for a review see the chapter by Cho, Lazarian \& Vishniac 
in this volume).

These results encourage us to believe that the many remaining
problems are ripe for further progress.  These problems 
range from the nature of dynamo
processes in stars, accretion disks, and galaxies, to the question
of how magnetic fields reconnect on dynamical time scales,
with apparent disregard for the constraint due to flux-freezing.
To be more precise, in the limit of
negligible resistivity, the magnetic field in a fluid medium 
is frozen, the sense that neither the magnetic flux threading a fluid element,
nor the field topology, can change.  Magnetic reconnection, the
exchange of partners between adjacent field lines violates the second
condition, while the generation of a large scale field through dynamo
action apparently violates the first.

Conventional {\it mean field dynamo} theory
(see \cite{m78,p79,kr80} for reviews)
allows a large-scale magnetic field to grow
exponentially from a seed field (see \cite{r88,l92})
at the expense of small-scale turbulent energy through
a process of spiral twisting and reconnection, illustrated
in Fig.~1.
\begin{figure}[t]
\begin{center}
\includegraphics[width=0.8\textwidth]{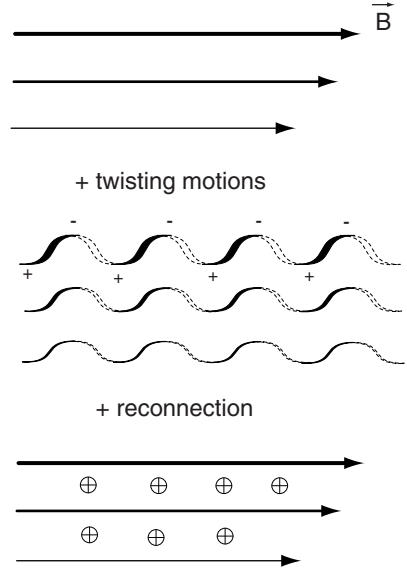}
\end{center}
  \caption{The mean-field dynamo in action.  Anisotropic turbulence
twists the field lines into spirals.  Reconnection restores the
original field lines, but a vertical gradient in the strength of the
spirals generates a net flux out of the page.}
\label{f1}
\end{figure}
This process starts with a set of large scale parallel field lines pointing
in some arbitrary direction. If the
underlying turbulence has a tendency to twist the field lines
into spirals with a preferred handedness (i.e. the velocity
field has some net helicity), then
reconnection on two dimensional surfaces between adjacent
spirals will produce a new field, at right angles to the
old one, provided that there is a systematic gradient in
the strength of the spirals.  The new field component is at
right angles to both the gradient and the old field component.
In a differentially rotating system, we can get a dynamo if
the original field direction is in the $\hat\phi$ direction,
and the dynamo process produces a radial field component.
Differential shearing of $B_r$ will then drive the azimuthal
field component, closing the cycle. This is the `$\alpha-\Omega$
dynamo'.  In the absence of global shear, we need
a second round of dynamo action, which gives an $\alpha^2$ dynamo.
This process can be given a systematic mathematical treatment by a suitable
choice of averaging procedures.

There are two ways in which this picture ignores, rather than
solves, the difficulties imposed by flux-freezing.  The more
obvious point is that adjacent spiral field lines are assumed
to reconnect quickly.  Without this assumption the field
will accumulate small scale tangled knots which will quickly
suppress dynamo action, and the large scale magnetic field
will saturate
far below equipartition with the surrounding turbulence.
Unfortunately, if reconnection happens at the rate allowed by the
generally accepted Sweet-Parker model \cite{p57,s58}, it is far too slow. 
However, there are observations which suggest that this
represents more of a challenge for theorists than a real
constraint on the evolution of magnetic fields.  If reconnection is
slow, turbulence would cause many magnetic
reversals per parsec within the interstellar medium.
Observations, on the contrary, show that magnetic field is
coherent over the scales of hundreds of parsecs. This
fact, as well as direct observations of large and small
scale Solar flares \cite{d95}, suggest that
the rate of reconnection is many orders of magnitude more rapid
than allowed by the Sweet-Parker model.  As this example shows,
the importance of reconnection in astrophysics is not limited to
understanding the dynamo process.  The process of reconnection
is an integral part of the transfer of magnetic energy to fluid
and particle motion in stellar coronae and in the interstellar medium.
More generally, it is impossible to claim that we understand MHD unless
we can predict whether crossing magnetic flux tubes will reconnect or bounce
from one another.  

A more subtle difficulty arises from the process by which straight
field lines are twisted into spirals.
This is intuitively appealing if we consider field lines as
isolated strings of infinitesimal radius.  More realistically,
the field occupies a non-zero volume.  Twisting a tube into a spiral
shape requires that we either allow the ends to slip, or allow
parts of the tube to twist in the opposite sense.  There is a geometrical
constraint which is ignored in the standard picture.  This
objection can be given a rigorous mathematical form,  the conservation
of magnetic helicity, which we will describe in \S 3.

How do numerical simulations of dynamo activity compare to mean-field
dynamo theory?
Computer simulations of dynamos can be divided into two classes.  There
are simulations in which some local instability (convection, the Balbus-Hawley
instability etc.) is allowed to operate, and there are simulations in
which the turbulence is driven externally, usually in such a way as to
guarantee the presence of a net fluid helicity.  The former simulations
are often successful at generating large scale magnetic fields whose
energy density is at least as great as the turbulent energy density
(e.g.\cite{bnst95,hb92,gr95}).
The latter are less successful, in the sense that the energy
density of the large scale magnetic field is often quite modest (e.g.
\cite{m81,b00}).  In particular there are
simulations (\cite{ch96,b01}) which produce dynamos in 
a computational box, with forced heliacal turbulence. These
dynamos show a steep inverse correlation
between the dynamo growth rate and the conductivity.  Naively
extrapolating to astrophysical regimes suggests that magnetic
dynamos driven by fluid helicity would take enormous amounts of
time to grow.  This conclusion is sharply at odds with evidence
for rapid and efficient stellar dynamos.

Here we discuss recent work on the problems of fast reconnection and
magnetic helicity conservation in astrophysical dynamos.  For
reconnection we concentrate
on a generic reconnection scheme that appeals to magnetic field
stochasticity as the critical property that accelerates reconnection
(\cite{lv99}, see \cite{lv00} for a review). 
Collisionless plasma effects which may also accelerate magnetic reconnection 
are addressed in the chapter by Bhattacharjee in this volume.  We will
typically assume that the evolution of the magnetic field is described by
the simplest form of the induction equation
\begin{equation}
\partial_t{\bf B}={\bf\nabla\times v\times B}+\eta\nabla^2{\bf B},
\label{induct}
\end{equation}
although we will make reference to work which includes more realistic
treatments of collisionless plasma effects.

\section{Rates of Magnetic Reconnection}

A simple dimensionless measure of the importance of resistivity,
$\eta$, in a conducting fluid is the Lundquist number $\equiv V_AL/\eta$, 
where $V_A\equiv B/(4\pi\rho)^{1/2}$ is the Alfv\'en velocity and $L$
is a typical scale of the system.  When fluid velocities are of order
the Alfv\'en speed, as is usual in astrophysics, this is a crude
estimate of the ratio of the first and second terms in equation (\ref{induct}).
Typically this number is very large
under most astrophysical circumstances, and flux freezing should
be a good approximation. More precisely, the coefficient of magnetic
field diffusivity in a fully ionized plasma is 
$\eta=c^2/(4\pi \sigma)=10^{13}T^{-3/2}$
cm$^2$ s$^{-1}$, where $\sigma=10^7 T^{3/2}$ s$^{-1}$ is the plasma
conductivity and $T$ is electron temperature. The characteristic time
for field diffusion through a plasma slab of size $L$ is
$L^2/\eta$, which is large for any ``astrophysical'' $L$.

What happens when magnetic field lines intersect? Do they deform
each other and bounce back or they do change their topology? This is
the central question of the theory of magnetic reconnection.
In fact, the whole dynamics of magnetized fluids and the back-reaction of
the magnetic field depends on the answer.

\subsection{The Sweet-Parker Scheme and its Modifications}

The literature on magnetic reconnection is rich and vast (see, for
example, \cite{pf00} and references therein). We start by discussing
a robust scheme proposed by Sweet and Parker \cite{p57,s58}. 
In this scheme oppositely directed magnetic fields are brought
into contact over a region of length $L_x$ (see Fig.~2). 
In general there will be a shared component, of the same order as
the reversed component.  However, this has only a minor effect
on our discussion.  The gradient
in the magnetic field is confined to the current sheet, a region
of vertical size $\Delta$, within which the magnetic
field evolves resistively. The velocity of reconnection, $V_r$,
is the speed with which magnetic field lines enter the current
sheet, and is roughly $\eta\approx V_r \Delta$.
Arbitrarily high values of $V_r$ can be achieved (transiently) by
decreasing $\Delta$. However, for sustained reconnection there is
an additional constraint imposed by mass conservation.
The plasma initially entrained on the magnetic field
lines must escape from the reconnection zone. In the Sweet-Parker
scheme this means a bulk outflow, parallel to the field lines, 
within the current sheet.  Since the mass enters along a zone of width
$L_x$, and is ejected within a zone of width $\Delta$, this implies
\begin{equation}
\rho V_{rec} L_x = \rho' V_A \Delta~~~,
\label{mass_con}
\end{equation}
where we have assumed that the outflow occurs at the Alfv\'en velocity.
This is actually an upper limit set by energy conservation.
If we ignore the effects of compressibility $\rho=\rho'$ and the
resulting reconnection velocity allowed by Ohmic diffusivity
and the mass constraint is 
\begin{equation}
V_{rec, sweet-parker}\approx V_A {\cal R}_L^{-1/2}, 
\end{equation}
where
${\cal R}_L$ is the Lundquist number using the current sheet {\it length}.
Depending on the specific astrophysical context, this gives a reconnection
speed which lies somewhere between $10^{-3}$ (stars) and $10^{-10}$ (the
galaxy) times $V_A$.

\begin{figure}
\begin{picture}(441,216)
\includegraphics{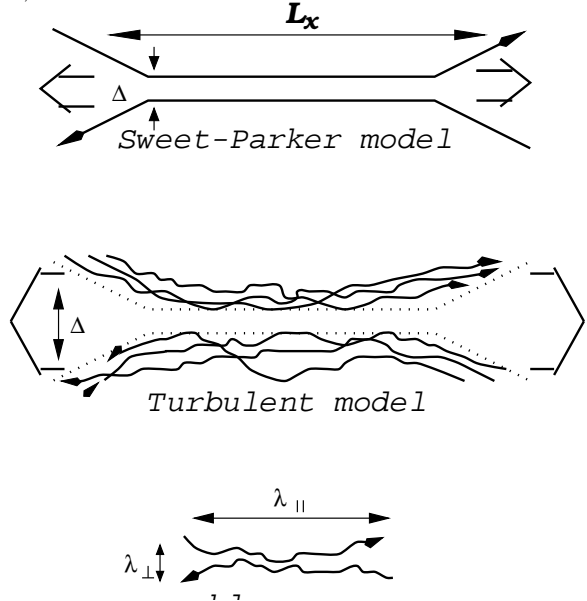}
\end{picture}
  \caption{Upper plot: Sweet-Parker scheme of reconnection. Middle plot:
illustration of stochastic reconnection that accounts for field line
noise.  Lower plot: a close-up of the contact region. 
Thick arrows depict outflows of plasma. From \cite{lv00}.}
\end{figure}

It is well known that using the
Sweet-Parker reconnection rate it is impossible to explain solar
flares.  For the reasons given in the introduction, it is also well
known that it is impossible to reconcile
dynamo theory with observations without some substantially faster
reconnection scheme.  Consequently, for forty years discussions of 
reconnection speeds have tended to focus on mechanisms that might
give reconnection speeds close to $V_A$, i.e. `fast' reconnection.  
In general, we can divide schemes for fast reconnection into those
which alter the microscopic resistivity, broadening the current sheet,
and those which change the global geometry, thereby reducing $L_x$.
Ultimately, a successful scheme should satisfy basic physical
constraints without requiring contrived geometries or boundary
conditions.  In the near term, we can gain some insight into the
likely nature of the solution by considering that reconnection is
not always fast.  Magnetic field lines in the solar corona and chromosphere
which could reach a lower energy configuration through reconnection do not 
always immediately do so.  Furthermore, a solution which relies 
entirely on collisionless
effects, for example, would imply that field lines do not reconnect in
dense environments, which would leave a major problem in understanding 
the nature of stellar dynamos. 

Attempts to accelerate Sweet-Parker reconnection are numerous.
We start by considering schemes to broaden the current sheet.
Anomalous resistivity is known to broaden current sheets in laboratory
plasmas.  It is present in the reconnection  layer when the
field gradient is so sharp that the electron drift velocity is
of the order of thermal velocity of ions $u=(kT/m)^{1/2}$\cite{p79}.
In other words, when $j>j_{cr}=Neu$.
If the current sheet has a width $\delta$ with a change in the
magnetic field $\Delta B$ then $4\pi j=c \Delta B/\delta$.
The effective resistivity increases nonlinearly as $j$ becomes greater
than $j_{cr}$, thereby broadening the current sheet.
We can find an upper limit to this effect by assuming that 
$j$ never gets very much larger than $j_{cr}$,
that is $\delta\approx\frac{c\Delta B}{4\pi N e u}$.
Expressing $\delta$ in terms of the ion cyclotron radius $r_c=(mu c)/(e B_{tot})$,
where $B_{tot}$ is the total magnetic field (including any shared
component) we find
\begin{equation}
\delta\approx r_c \left(\frac{V_A}{u}\right)^2 \frac{\Delta B}{B_{tot}}~~~,
\label{deltrc}
\end{equation}
which agrees with \cite{p79} up to the factor
$\Delta B/B_{tot}$, which equals 1 in that treatment. Combining
(\ref{mass_con}) and (\ref{deltrc}) one gets \cite{lv99}
\begin{equation}
V_{rec, anomalous}\approx V_A \frac{r_c}{L_x} 
\left(\frac{V_A}{u}\right)^2 \frac{\Delta B}{B_{tot}}~~~.
\label{rec:anom}
\end{equation}
Equation (\ref{rec:anom}) shows that the enhanced
reconnection velocity is still much less than
the Alfven velocity if
$L_x$ is much greater than the ion Larmor (cyclotron) radius.
In general, ``anomalous reconnection'' is important when
the thickness of the reconnection layer in the Sweet-Parker
reconnection scheme is less than $\delta$. However, for typical interstellar magnetic
fields the Larmor radius $r_c$ is $\sim 10^7$~cm and anomalous
effects are negligible.

Tearing modes are a robust instability connected to the
appearance of narrow current sheets \cite{fkr63}.
The resulting turbulence
will broaden the reconnection layer and enhance the reconnection
speed.  Here we give an estimate of this effect and show that
while it represents a significant enhancement of Sweet-Parker
reconnection of laminar fields, it leaves reconnection slow.
One difficulty with many earlier
studies of reconnection in the presence the tearing modes
stemmed from the idealized two dimensional geometry
assumed for reconnection.  In two dimensions tearing
modes evolve via
a stagnating non-linear stage related to the formation of magnetic
islands.  This leads to a turbulent reconnection zone \cite{ml85},
but the current sheet remains narrow and its effects on the 
overall reconnection speed are unclear.
This nonlinear stagnation stage does not emerge when realistic 
three dimensional configurations are considered \cite{lv99}.
In any realistic circumstances field lines are not
exactly antiparallel.
Consequently, we expect that instead of islands one finds nonlinear
Alfv\'en waves in three dimensional reconnection
layers.  The tearing instability proceeds with growth rates
determined by the linear growth phase while the resulting magnetic structures
propagate out of the reconnection region at the Alfv\'en speed.

The dominant mode will be the longest wavelength mode,
whose growth rate will be 
\begin{equation}
\gamma\approx {\eta\over\Delta^2}\left({V_A\lambda_{\|}\over\eta}\right)^{2/5}.
\end{equation} 
The transverse spreading of the plasma in the reconnection layer will
start to stabilize this mode when its growth
rate is comparable to the transverse shear $V_A/\lambda_{\|}$ \cite{bss79}.
At this point we have $V_{rec,local}\approx\gamma\Delta$ and \cite{lv99} 
\begin{equation}
V_{rec, tearing}=V_A\left({\eta\over V_A L_x}\right)^{3/10},
\label{tearing}
\end{equation}
which is substantially faster than the Sweet-Parker rate, but still
very slow in any astrophysical context. Note that unlike anomalous
effects, tearing modes do not require any special conditions and therefore
should constitute a generic scheme of reconnection.

Finally, we note that there is a longstanding, but controversial suggestion,
that ions tend to scatter about once per cyclotron period, `Bohm diffusion'
\cite{b49}. 
Even if this is correct, the effective diffusivity of magnetic field lines
would still be only $\eta_{\rm Bohm}\sim V_A r_c$.  While this would
be a large increase over Ohmic resistivity,
it produces fast reconnection, of order $V_A$, only if $r_c\sim L_x$.
It therefore fails as an explanation for fast reconnection for the
same reason that anomalous resistivity does.

\subsection{X-point Reconnection}

The failure to find fast reconnection speeds through current sheet 
broadening has stimulated interest in
fast reconnection through radically different global 
geometries.  Petschek \cite{p64} conjectured that reconnecting magnetic
fields would tend to form structures whose typical size in
all directions is determined by the resistivity (`X-point' reconnection).
This results in
a reconnection speed of order $V_A/\ln {\cal R}_L$.  However,
attempts to produce such structures in numerical simulations
of reconnection have been disappointing.  Typically
the X-point region collapses towards the
Sweet-Parker geometry as the Lundquist number becomes
large \cite{b84,b86,b96,wmb96,wb96}.\footnote{Recent plasma reconnection experiments 
\cite{y00} do not support Petschek scheme either.}
One way to understand this collapse is to consider perturbations
of the original X-point geometry.
In order to maintain this geometry
shocks are required in the original
(Petschek) version of this model.  These shocks are, in turn, supported
by the flows driven by fast reconnection, and fade if $L_x$ increases.
Naturally, the dynamical range for which the existence of such shocks
is possible depends on the Lundquist number and shrinks when
fluid conductivity increases.  The apparent conclusion is that, 
at least in the collisional regime,
reconnection occurs through narrow current sheets. 

One may invoke collisionless plasma effects
to stabilize the X-point reconnection (for collisionless plasma).
For instance, a number of authors \cite{sddb98,sd98,sdrd99}
have reported that in a two fluid treatment of magnetic reconnection,
a standing whistler mode can stabilize an X-point with a scale comparable
to the ion plasma skin depth, $c/\omega_{pi}\sim (V_A/c_s)r_L$.
The resulting reconnection speed is a large fraction of $V_A$, and
apparently 
independent of $L_x$, which would suggest that something like
Petschek reconnection emerges in the collisionless regime. 
This possibility
is discussed at length in the chapter by Bhattacharjee (this volume).  
However,
these studies have not yet demonstrated the possibility of fast reconnection
for generic field geometries, since they assume that there are no
bulk forces acting to produce a large scale current
sheet. Similarly, those studies do not account for fluid turbulence. 
Magnetic fields embedded in a turbulent fluid will give
fluctuating boundary conditions for the current sheets.
On the other hand,  boundary conditions need to be fine tuned for a
Petschek reconnection scheme \cite{pf00}.

Finally, we note that 
a number of researchers have claimed that turbulence may accelerate
reconnection (for example, \cite{s88}, where tearing modes are
used as the source of the turbulence).  The general idea
is that turbulent motions can provide an effect transport
coefficient $\sim \langle v^2\rangle\tau$ \cite{p79}.  However,
a closer examination of this process has convincingly demonstrated
that an unrealistic amount of energy is required to mix field
lines unless they are almost exactly anti-parallel \cite{p92}. 
In the next section we will discuss
a mechanism that, when it works,
should produce reconnection under a broad range of field geometries,
without regard to the particle collision rate.

\subsection{Stochastic Reconnection}

Two idealizations were used in the preceding discussion. First,
we considered reconnection in only two dimensions.  Second, we assumed
that the magnetized plasma has laminar field lines. 
The Sweet-Parker scheme can
easily be extended into three dimensions, in the sense that one can take
a cross-section of the reconnection region such that the shared
component of the two magnetic fields is perpendicular to the
cross-section.  
In terms of the mathematics nothing changes, but the outflow velocity
becomes a fraction of the total $V_A$ and the shared component of the
magnetic field will have to be ejected together with the plasma. 
This result has motivated
researchers to do most of their calculations in 2D, which has obvious
advantages for both analytical and numerical investigations.

However, physics in two and three dimensions are very different.
This is true, for example, in hydrodynamic turbulence, partly
because lines of vorticity have different dynamics when they
are free to move around one another.  Similarly, the ability
of magnetic field lines to move past one another in three dimensions
dramtically alters the topolical constraints on their dynamics.
In \cite{lv99} we
considered three dimensional
reconnection in a turbulent magnetized fluid and showed that reconnection
is fast. This result cannot be obtained by considering two dimensional
turbulent reconnection (cf. \cite{ml85}).  This point has been
the source of significant confusion.  Turbulent reconnection has
usually been used to refer to reconnection driven by the
turbulent transport of magnetic flux, as discussed in the previous
subsection.  In other words, one looks for a net flux transport term,
operating on microscales, that is proportional to magnetic field
gradients and has a coefficient which is independent of the resistivity.
This process was recently examined, and severely criticized, in \cite{kd01}, 
under the mistaken impression that it the critical physical
process in stochastic reconnection.  Instead, stochastic reconnection 
is a geometric effect arising from the appearance of stochastic
field line wandering in three dimensions, which gives rise to a broad
outflow from the current sheet, but has little effect on the current
sheet structure.  Below we
briefly discuss the idea of stochastic reconnection, while the
full treatment of the problem is given in \cite{lv99}.

MHD turbulence guarantees the presence of a stochastic field component,
although its amplitude and structure clearly depends on the 
amplitude and the turbulence driving mechanism.  Our {\it model}
of the field line stochasticity also depends on our ability to model
generic MHD turbulence.
We consider the case in which there exists a large scale,
well-ordered magnetic field, of the kind that is normally used as
a starting point for discussions of reconnection.  This field may,
or may not, be ordered on the largest conceivable scales.  However,
we will consider scales smaller than the typical radius of curvature
of the magnetic field lines, or alternatively, scales below the peak
in the power spectrum of the magnetic field, so that the direction
of the unperturbed magnetic field is a reasonably well defined concept.
In addition, we expect that the field has some small scale `wandering' of
the field lines.  On any given scale the typical angle by which field
lines differ from their neighbors is $\phi\ll1$, and this angle persists
for a distance along the field lines $\lambda_{\|}$ with
a correlation distance $\lambda_{\perp}$ across field lines (see Fig.~2).

The modification of the mass conservation constraint in the presence of
a stochastic magnetic field component 
is self-evident. Instead of being squeezed from a layer whose
width is determined by Ohmic diffusion, the plasma may diffuse
through a much broader layer, $L_y\sim \langle y^2\rangle^{1/2}$ 
determined by the diffusion of magnetic field lines.
(Here `$y$' is the axis perpendicular to the mean field direction. 
See Fig.~2.)
This suggests
an upper limit on the reconnection speed of 
$\sim V_A (\langle y^2\rangle^{1/2}/L_x)$. 
This will be the actual speed of reconnection if
the progress of reconnection in the current sheet itself does not
impose a smaller limit. The value of
$\langle y^2\rangle^{1/2}$ can be determined once a particular model
of turbulence is adopted, but it is obvious from the very beginning
that this value is determined by field wandering rather than Ohmic
diffusion, as in the Sweet-Parker model. 

What about limits on the speed of reconnection that arise from
considering the structure of the current sheet?
In the presence of a stochastic field component, magnetic reconnection
dissipates field lines not over their  entire length $\sim L_x$ but only over
a scale $\lambda_{\|}\ll L_x$ (see Fig.~2), which
is the scale over which magnetic field line deviates from its original
direction by the thickness of the Ohmic diffusion layer $\lambda_{\perp}^{-1}
\approx \eta/V_{rec, local}$. If the angle $\phi$ of field deviation
did not depend on the scale, the local
reconnection velocity would be $\sim V_A \phi$, independent of 
resistivity. However, for any realistic model of MHD turbulence, $\phi$ 
($=\lambda_{\perp}/\lambda_{\|}$,
does depend on scale. Consequently, the {\it local}
reconnection speed $V_{rec,local}$ is given by the usual Sweet-Parker formula
but with $\lambda_{\|}$ instead of $L_x$, i.e. $V_{rec, local}\approx V_A 
(V_A\lambda_{\|}/\eta)^{-1/2}$.  Also, it is apparent
from Fig.~2 that $\sim L_x/\lambda_{\|}$ magnetic field 
lines will undergo reconnection simultaneously (compared to a one by one
line reconnection process for
the Sweet-Parker scheme). Therefore the overall reconnection rate
may be as large as
$V_{rec, global}\approx V_A (L_x/\lambda_{\|})(V_A\lambda_{\|}/\eta)^{-1/2}$.
Whether or not this limit is important depends on
the value of $\lambda_{\|}$.  

The relevant values of $\lambda_{\|}$ and $\langle y^2\rangle^{1/2}$
depend on the magnetic field statistics. This
calculation was performed in \cite{lv99} using the Goldreich-Sridhar
model \cite{gs95}
of MHD turbulence, the Kraichnan model (\cite{i63,k65})
and for MHD turbulence with an arbitrary spectrum (limited only
some basic physical constraints and which is in rough agreement with
observations \cite{a95,lp00,sl02}).
In all the cases the upper limit on $V_{rec,global}$ was greater
than $V_A$, so that the diffusive wandering of field lines imposed
the relevant limit on reconnection speeds.
Among these, the Goldreich-Sridhar model provides the best fit to 
observations (e.g. \cite{a95,sl02})  and simulations \cite{cva,cvb}.  
In this case the 
reconnection speed was 
\begin{equation}
V_{rec, up}=V_A \min\left[\left({L_x\over l}\right)^{\frac{1}{2}},
\left({l\over L_x}\right)^{\frac{1}{2}}\right]
\left({v_l\over V_A}\right)^{2},
\label{main}
\end{equation}
where $l$ and $v_l$ are the energy injection scale and
turbulent velocity at this scale respectively.  We stress that
the use of MHD turbulence models here is solely for the purpose
of providing a well-defined model of field line stochasticity.
The dynamics of the turbulent cascade are largely irrelevant and
any process which leads to small scale field line stochasticity
(e.g. footpoint motions for solar field lines) is a possible
cause of fast reconnection.

In \cite{lv99} we also considered other processes that can impede
reconnection and find that they are less restrictive. For
instance, the tangle of reconnection field lines crossing the
current sheet will need to reconnect repeatedly before individual
flux elements can leave the current sheet behind.  The rate at which
this occurs can be estimated by assuming that it constitutes the
real bottleneck in reconnection events, and then analyzing each
flux element reconnection as part of a self-similar system of
such events.  This turns out to limit reconnection to speeds less
than $V_A$, which is obviously true regardless.  As the result 
equation (\ref{main}) is not only an
upper limit on the reconnection speed, but is the best estimate of 
its value.

Naturally, when turbulence is negligible, i.e. $v_l\rightarrow 0$, the
field line wandering is limited to the Sweet-Parker current sheet
and the Sweet-Parker reconnection scheme takes over. However, in practice
this requires an artificially low level of turbulence that
should not be expected in realistic astrophysical environments.
Moreover, the release of energy due to reconnection, at any speed,
will contribute to the turbulent cascade of energy and help drive
the reconnection speed upward.  This may be relevant to the slow
onset, and rapid acceleration, of the reconnection process in 
solar flares. 

We stress that the enhanced reconnection efficiency in turbulent
fluids is only present if 3D reconnection is considered. In
this case ohmic diffusivity fails to constrain the reconnection process
as many
field lines simultaneously enter the reconnection region. The
number of lines that can do this increases with the decrease of
resistivity and this increase overcomes the slow rates of
reconnection of individual field lines. It is impossible to achieve
a similar enhancement in 2D (see \cite{z98}) since field lines
can not cross each other.

There is a limited analogy one can draw between the enhancement of
reconnection speeds in X-point models and increased rate of reconnection
due to field line stochasticity.  In both cases one gets a boost from
a reduced parallel length scale.  In the case of X-point models this
effect is, usually by design, enormous since $L_x\rightarrow \Delta$.
Stochastic reconnection depends on a relatively modest enhancement,
since $L_x\rightarrow \lambda_{\|}(\Delta)\gg\Delta$.  The bulk of the
effect comes from the simultaneous reconnection of many independent
flux elements, and the steady diffusion of the ejected plasma away
from the current sheet.  The main problem with X-point reconnection
models, their tendency to collapse to narrow current sheets, is absent
in stochastic reconnection, since in the latter case the current sheets
stay narrow, and the diverging field lines are separated by other field
lines, rather than by unmagnetized plasma.

A more subtle difficulty arises from our prescription for the structure
of the stochastic field near the current sheet.  We have assumed that
we can apply the statistically homogeneous prescription for field line
perturbations in a turbulent medium near planes where there is a dramatic
change in the structure of the large scale magnetic field.  This is
not obvious.  It may be that the presence of a strong shear in the field
acts as a kind of internal surface, producing an altered, and perhaps
greatly reduced, level of stochasticity.  This kind of internal
`shadowing' does not appear in current simulations, but there has been
little attempt to look for it, and the issue can only be resolved
when detailed numerical simulations of stochastic reconnection are 
performed.  Similarly, one may wonder if the systematic ejection of
plasma along the field lines might modify their topological connections.
In this case it seems more plausible to suppose that this would lead
to an increase in the diffusion rate, rather than a decrease, but
again no simulations of this process are available.

\subsection{Reconnection in Partially Ionized Gas}

A substantial fraction of the ISM in our galaxy is partially ionized,
as well as photospheres of most stars.  This motivates studies
of the effect of neutrals on reconnection and MHD turbulence.
The role of ion-neutral collisions is not trivial. On one hand, 
neutral particles tend to have a substantially longer mean free
path, so that drag between the neutrals and ions 
may truncate the turbulent cascade at a relatively large scale.  
On the other hand, the ability of neutrals to diffuse perpendicular
to magnetic field lines enhances reconnection rates, at least in the
Sweet-Parker model.

Reconnection in partially ionized gases has been studied by
various authors (\cite{nma92,zb97,vl99}) in the context
of the Sweet-Parker reconnection model.  Our comments
here are based on \cite{vl99}
where we studied
the diffusion of neutrals away from the reconnection zone. 
In general, in a partially ionized gas the reconnection zone consists of two
distinct regions. A broad region, which width is determined by the ambipolar diffusivity,
$\eta_{ambi}\approx V_{A}^2/t_{ni}$  where $t_{ni}$
is the neutral-ion collision rate, and a
narrow region whose width is determined by the Ohmic diffusivity.  
Magnetic reconnection takes place in the narrow region, while the
broader region allows a more efficient ejection of matter.

If the recombination time is short, then ions and neutrals are largely
interchangeable and the reconnection speed is \cite{vl99}
\begin{equation}
V_{rec}\approx V_A \left({V_A t_{in}\over L_x}\right)^{1/2}.
\end{equation}
This is faster than the Sweet-Parker rate, but not fast in the sense of
allowing reconnection speeds close to $V_A$.  In practice, even this
rate is typically unachievable.  Under typical interstellar conditions
the reconnection speed is limited by the recombination rate.  That is,
the rate at which ions recombine and leave the resistive region determines
the speed of the whole process.
Consequently,
the ambipolar reconnection rates obtained in \cite{vl99}
are insufficient either
for fast dynamo models or for the ejection of magnetic flux prior
to star formation.  In fact, the increase in the reconnection speed
stems entirely from the
compression of ions in the current sheet, with the consequent enhancement of
both recombination\footnote{In the model \cite{vl99} it is assumed that the
ionization is due to cosmic rays. In the case of photoionization of the
heavy species, e.g. carbon, the recombination and therefore the reconnection
rates are lower.} and ohmic dissipation.  This effect is small
unless the reconnecting magnetic field lines are almost exactly
anti-parallel.  As above, we expect that including the effects
of anomalous resistivity and tearing modes may enhance reconnection
speeds appreciably, but not to the extent of producing fast
reconnection. 

None of this work included the effects of field line stochasticity, which is critical
for producing fast reconnection in ionized plasmas.  
We expect that in this case also the presence of turbulence will
lead to substantially higher reconnection speeds.  However,
whether
or not this produces fast reconnection must depend on the nature
of the turbulent cascade in a partially ionized gas.
Recent work, which is discussed in detail in the chapter by
Cho, Lazarian \& Vishniac in this volume, show that the magnetic field in a 
partially ionized gas has a much more complex structure than it is usually assumed.
In fact, in \cite{clv02} we reported a new regime of MHD
turbulence which is characterized by the existence of intermittent
magnetic structures below the viscous cutoff scale.  The root mean
square perturbed magnetic field strength in these structures does not
drop at smaller scales.  However, the {\it curvature} scale
(and therefore the divergence rate) for these structures does not
decrease significantly as their perpendicular scale decreases. 
At sufficiently
small scales the ions and neutrals will decouple, and a turbulent cascade,
extending down close to resistive scales but involving only ions will appear.

The existence of strong magnetic field structures on small scales, and
the reappearance of a strong turbulent cascade at very small scales, should 
lead to fast reconnection speeds through stochastic reconnection.  
However, it remains to be seen whether or not the intermediate scales,
characterized by weak divergence of field lines, will impose a significant
bottleneck on the reconnection plasma outflow.  If it does, then the
implication is that interstellar clouds with small ionized fractions may
not allow fast reconnection. This conclusion would not pose any problems
with galactic dynamo, but may be extremely important for other
essential processes, e.g. star formation.  This issue is examined further
in \cite{lvc02}.

\section{The Dynamo Process}

\subsection{Conventional Theory and its Problems}

We start this section by briefly reviewing the standard approach to dynamo
theory, and discussing various objections to it.  Some of these objections
center around the speed of reconnection, and 
can be safely ignored if reconnection is fast in a turbulent environment.
In fact, since stochastic reconnection depends on small scale structure
in the magnetic field, the claim that small scale structure tends to
accumulate energy faster than the large scale field \cite{ka92} can 
be seen as self-limiting.  A disproportionate growth in power on small
scales will only continue until the reconnection speed is boosted to
large fraction of $V_A$.
However, as we have already mentioned, some objections to dynamo
theory are more subtle and require
substantial modification to mean-field dynamo theory.

The usual approach to the dynamo problem is to take equation (\ref{induct}),
set $\eta=0$, and divide the velocity field into small scale turbulence
and some large scale rotational motion.  In
order to follow the evolution of the large scale magnetic
field, we write
\begin{equation}
{\bf B}\equiv \langle{\bf B}\rangle +{\bf b}.
\end{equation}
The brackets here denote averaging over scales somewhat
larger than the turbulent eddy size.  In other words, they
indicate a smoothing process which averages out all small
scale features.  The field $\langle{\bf B}\rangle$ is the `mean field'.
The dynamo process can be written in
mathematical terms by approximating the
evolution of the small scale field component, ${\bf b}$,
as
\begin{equation}
\partial_t{\bf b}\approx {\bf\nabla\times v\times}
\langle {\bf B}\rangle,
\end{equation}
and substituting the result into the evolution equation for the
large scale field,
\begin{equation}
\partial_t\langle{\bf B}\rangle ={\bf\nabla\times}
\langle {\bf v\times b}\rangle.
\label{ls}
\end{equation}
In a turbulent, incompressible and homogeneous plasma this implies
\begin{equation}
\partial_t\langle{\bf B}\rangle=
{\bf\nabla\times}({\bf \alpha\cdot B})
+{\bf\nabla\cdot}({\bf D_T\cdot\nabla})\langle{\bf B}\rangle.
\label{dyn1}
\end{equation}
Here $\alpha$, the kinetic helicity, and ${\bf D_T}$, the turbulent
diffusion tensor, are dyads given by
\begin{equation}
\alpha_{il}\equiv\epsilon_{ijk}\langle v_j\partial_l v_k\rangle\tau_c,
\label{khel}
\end{equation}
and
\begin{equation}
D_{T,ij}\equiv\langle v_iv_j\rangle\tau_c,
\end{equation}
where $\tau_c$ is the eddy correlation time.  The component of the
electromotive force along the large scale field direction,
$\langle{\bf\hat B}\rangle\cdot\langle{\bf v\times b}\rangle$,
is the piece that can drive an increase in the large scale
magnetic field.  (The component perpendicular to $\langle{\bf B}\rangle$ 
gives an effective large scale field velocity, that is, it affects
the transport of the field rather than its generation.)
The trace of $\alpha$ divided by $\tau_c$
is what is usually referred to as the kinetic helicity, and it
is often assumed for convenience that ${\bf \alpha}$ is a scalar times the identity 
matrix.  In symmetric turbulence $\alpha$ vanishes, 
but ${\bf D_T}$ does not.
In fact, since a successful dynamo requires non-vanishing
diagonal components for $\alpha$, we can see from
this expression that a successful dynamo should require symmetry
breaking along all three principal axes.

The appearance of $D_T$ in equation (\ref{dyn1}) would seem to vindicate
the use of turbulent diffusion in astrophysical MHD.  There are two
reasons why this is not quite right.  First, fast reconnection is 
implicit in this kind of averaging argument.  Rather than appealing to
turbulent diffusion as an explanation for fast reconnection, we are
actually using our understanding of fast reconnection to explain
diffusion.  The second point is less formal and more important. 
Equation (\ref{dyn1}) is not a realistic description of the evolution
of $\langle{\bf B}\rangle$.  As noted in \S 1, twisting magnetic 
field lines into spirals is not easily accomplished, and numerical
simulations do not support the use of equation (\ref{dyn1}).

\subsection{Magnetic Helicity Conservation Constraint}

The fundamental problem is that there is an important mathematical
constraint that follows from equation (\ref{induct}), which is not
respected by equation (\ref{dyn1}).  The magnetic
helicity, defined as $H\equiv {\bf A}\cdot{\bf B}$
evolves according to 
\begin{equation}
\partial_t H=-{\bf\nabla\cdot}\left[{\bf A\times}({\bf v\times B}+
{\bf\nabla\Phi})\right]-\eta{\bf B\cdot\nabla\times B},
\label{hel}
\end{equation}
where $\Phi$ is an arbitrary function of space and time.
For the Coulomb gauge, which turns out to be a convenient choice, we require
\begin{equation}
\nabla^2\Phi={\bf\nabla\cdot}({\bf v\times B}).
\end{equation}
In the limit of vanishing resistivity, this not only implies that the
volume integrated magnetic helicity vanishes, it also implies that the
magnetic helicity of any individual flux tube is separately
conserved \cite{t74}.

For a non-zero, but very small, $\eta$, we can transfer magnetic
helicity from one flux tube to another.  However,
since $H$ is of order $L B^2$, where $L$ is a characteristic
scale of the field, it takes less energy to hold magnetic helicity
on large scales than on eddy scales, and a divergent amount on
infinitesimal scales.  Consequently, in the limit of vanishing resistivity
the resistive term in equation (\ref{hel}) does not affect the
global conservation of helicity, even in the presence of fast reconnection, 
as long as reconnection only occurs in an infinitesimal fraction of the
plasma volume.  On the other hand, the conservation of magnetic helicity
for individual flux tubes is completely lost.  The implication is
that global magnetic helicity conservation is a good approximation
for laboratory plasmas,
a point that was originally stressed by Taylor \cite{t74},
and an even better one for astrophysical systems.

How does this affect dynamo theory?  The large scale distribution
of magnetic helicity can be divided into a piece carried by large
scale magnetic structures and a piece carried by small scale structures,
or
\begin{equation}
\langle H\rangle=\langle{\bf A}\rangle\cdot\langle{\bf B}\rangle
+\langle{\bf a}\cdot{\bf b}\rangle.
\end{equation}
Henceforth we will use $h\equiv\langle{\bf a}\cdot{\bf b}\rangle$.
The evolution of the first piece, in a perfectly conducting fluid, is
\begin{equation}
\partial_t(\langle{\bf A}\rangle\cdot\langle{\bf B}\rangle)
=2\langle{\bf B}\rangle\cdot\langle{\bf v\times b}\rangle
-{\bf\nabla\cdot}
\left[\langle{\bf A}\rangle{\bf\times}(\langle{\bf v\times b}\rangle+
{\bf\nabla}\langle\Phi\rangle)\right].
\label{htrans}
\end{equation}
The second term on the right hand side is the magnetic helicity transport
driven by mean-field terms.  The first represents the exchange of magnetic
helicity between large and small scales.  This term is proportional to the
component of the electromotive force which drives the dynamo process.
In other words, the generation of a large scale magnetic field is a direct
consequence of the transfer of magnetic helicity between large and small
scales.

The point that MHD turbulence transfers magnetic helicity to the largest
available scales, even if that scale is much larger than any eddy scale,
is well known \cite{f75,smg94,smo95}.
We can estimate the rate at which $h$ is transferred to large scale magnetic
field structures by considering its role in biasing the value of the
electromotive force \cite{vc01}. The inverse cascade rate is
\begin{equation}
\tau_{cascade}^{-1}\sim {V_A^2\over\langle v^2\rangle} \tau_c^{-1}.
\end{equation}
For a large scale magnetic field in equipartition with the
turbulent cascade this implies that magnetic helicity is transferred
to the large scale field in one eddy turn over time.  This suggests that
unless the large scale field is very weak it is reasonable to take
\begin{equation}
H\approx\langle{\bf A}\rangle\cdot\langle{\bf B}\rangle.
\label{hbig}
\end{equation}
Then combining equations (\ref{htrans}) and (\ref{hbig}) we see that
\begin{equation}
2\langle{\bf B}\rangle{\bf \cdot}\langle{\bf v\times b}\rangle
=-{\bf\nabla\cdot}\left[\langle{\bf a\times}({\bf v\times B}+{\bf\nabla}\phi)
\rangle\right]\equiv-{\bf\nabla\cdot J}_H,
\label{div}
\end{equation}
where $J_H$ is defined as the magnetic helicity current carried by small
scale structures, or the anomalous magnetic helicity current.  
That is, the component of the electromotive force parallel to the large
scale magnetic field is given by the divergence of the magnetic helicity
current carried by eddy scale structures.  If $J_H\approx 0$, then
it follows from equation (\ref{div}) that mean-field dynamos are
impossible.  This argument was advanced by Gruzinov and Diamond
\cite{gd94,gd96} who pointed out that magnetic helicity conservation
combined with the assumption of stationary statistics for small
scale structure implied almost complete suppression of the kinematic
dynamo.  We note also that the form of the
parallel component of the electromotive force given in equation (\ref{div})
has been suggested
before \cite{bh86}, although the interpretation that the 
relevant current is a magnetic helicity current appeared somewhat
later \cite{j99,kmrs00}.  Here we will follow the treatment in
\cite{vc01}, where the magnetic helicity current was derived for the
first time for homogeneous turbulence.

Equations (\ref{ls}) and (\ref{div}) yield
\begin{equation}
\partial_t\langle{\bf B}\rangle={\bf\nabla\times}
\left[{-\langle{\bf B}\rangle\over2\langle B\rangle^2}{\bf\nabla\cdot J}_H
+\langle{\bf v\times b}\rangle_{\perp}\right],
\end{equation}
where the second term on the right hand side is the component of the
electromotive force perpendicular to the large scale field direction.
Evaluating ${\bf J}_H$ is necessary
to understand the dynamo process.  By contrast, attempts to estimate the
kinetic helicity only tell us about the dynamo process when the large
scale magnetic field is so weak that the transfer of magnetic helicity
between scales is unaffected by the extremely limited capacity of the
turbulent eddies to store magnetic helicity.

The most direct way to estimate the anomalous magnetic helicity current
is to write ${\bf a}$ in terms of the action of the turbulent velocity
field on the large scale magnetic field, or
\begin{equation}
{\bf a}\approx ({\bf v\times B}-{\bf\nabla}\phi)\tau_c,
\label{step}
\end{equation}
where
\begin{equation}
{\bf\nabla}^2\phi={\bf\nabla\cdot}({\bf v\times B}).
\end{equation}
If we substitute this into the definition of the magnetic helicity current
we find, after some manipulation, that
\begin{equation}
{\bf J}_H=-\tau_c \int{d^3{\bf r}\over4\pi r}\epsilon_{lmn}\langle B_k\rangle
\langle B_l\rangle
\langle \partial_k\partial_m\langle v_i({\bf x})v_n({\bf x}+{\bf r})
\rangle.
\label{jh1}
\end{equation}
We see that ${\bf J}_H$ is parity-invariant, unlike $\alpha$.  In completely
isotropic turbulence it will also vanish, but the degree of symmetry breaking
necessary for a dynamo effect is smaller than in the conventional picture.
There will also be contributions to $J_H$ driven by the effects of
background structure, but for strongly rotating systems these will be smaller 
than the expression given here.

Equation (\ref{jh1}) is not a particularly enlightening expression, but we can
gain somewhat more insight by rewriting it as
\begin{equation}
{\bf J}_H\approx-\lambda_c^2\tau_c\langle\langle{\bf B}\rangle{\bf\cdot\omega}
(\langle{\bf B}\rangle{\bf\cdot\nabla}){\bf v}\rangle,
\label{jh2}
\end{equation}
where $\lambda_c$ is some suitably averaged eddy size and
${\bf\omega}\equiv {\bf\nabla\times v}$ is the fluid vorticity.
This corresponds to twisting a field line in both directions, but
then systematically moving right and left handed spiral segments in
opposite directions.

In this model, we generate left (or right) handed spirals by separating
segments of the same field line with different helicities, moving them
in opposite directions, and then reconnecting them (along two dimensional)
surfaces, into new field lines.  If the flow of magnetic helicity has
a non-zero divergence, then the new field lines will have a preferred
sense of twisting, and the first step in the usual scheme for the dynamo
process will have been completed without violating magnetic helicity
conservation.  We illustrate this modified version of mean-field dynamo
action in Fig.~3.
\begin{figure}[t]
\begin{center}
\includegraphics[width=0.8\textwidth]{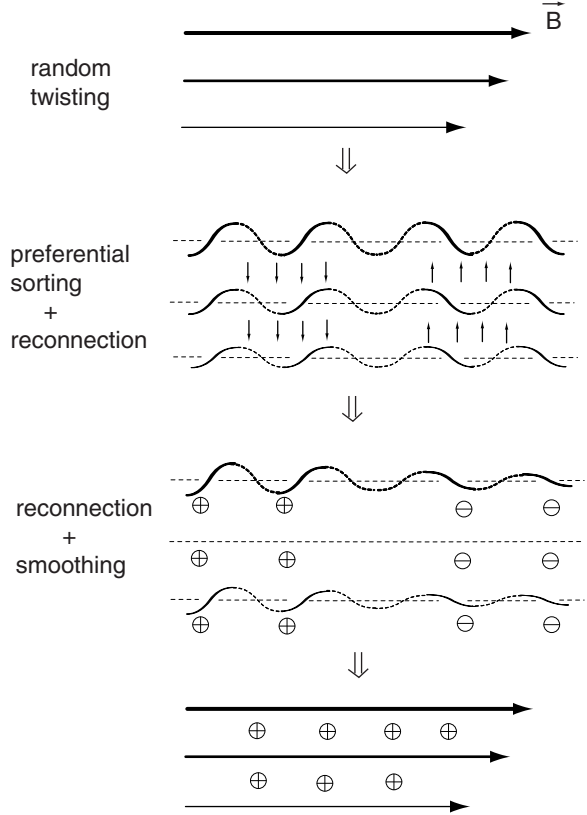}
\end{center}
  \caption{A new version of the mean-field dynamo.  Turbulence
twists the field lines into spirals, with each field line accumulating
regions of right and left handed twisting.  The turbulence is biased
so that left-handed segments move down and right-handed segments move
up.  Reconnection, and a gradient in the strength of the spirals gives
us new field lines with a net left-handed twist. The last step involves
averaging over each field line, which is trivial but not part of the
original picture, and reconnection to produce a new orthogonal field
component.}
\label{f3}
\end{figure}

In some sense equation (\ref{div}) gives the 
{\it minimal} change in dynamo theory which respects
conservation of magnetic helicity, since it leaves
leaving ${\cal E}_{mf\perp}$ unchanged.  That is, we assume that 
the backreaction
from small scales affects only the component of the electromotive
force along the direction of the large scale mean field.
While this may seem unduly optimistic, we note that in the absence
of any other large scale vector quantity, symmetry considerations
alone should be sufficient grounds for this assumption.
However, we are often concerned with circumstances where other large
scale vectors are present, for example systems with differential
rotation.  In this case we are not guaranteed that 
$\langle {\bf v\times b}\rangle_{\perp}$ is unaffected by the inverse
cascade.  Obviously this is an important direction for future work.

There is another way in which equation (\ref{jh1}) may fail to give
the full magnetic helicity current.  Equation (\ref{step}) takes into account
the perturbations in the large scale magnetic field driven by small
scale velocties.  However, fluctuations in the magnetic field also drive
the velocity field.  In conventional mean-field dynamo
theory these considerations lead to replacing equation  
(\ref{khel}) with
\begin{equation}
\alpha_{il}\equiv\epsilon_{ijk}\left[\langle v_j\partial_l v_k\rangle
-{\langle b_j\partial_l b_k\over 4\pi\rho}\rangle\right]
\tau_c.
\label{khel2}
\end{equation}
In our case equation (\ref{jh1}) can be replaced by a symmetrized
version, that is
\begin{equation}
{\bf J}_H=-\tau_c \int{d^3{\bf r}\over4\pi r}\epsilon_{lmn}\langle B_k\rangle
\langle B_l\langle\partial_k\partial_m\left[\langle v_i({\bf x})v_n({\bf x}+{\bf r})
-{b_i({\bf x})b_n({\bf x}+{\bf r})\over4\pi\rho}\rangle\right].
\label{jh3}
\end{equation}
This form fails to take into account the effects of shear from a large scale
velocity field (like differential rotation).  
Given perfect symmetry between the dynamics of the velocity and magnetic
fields, this term will be zero, regardless of any spatial symmetry-breaking effects.
MHD turbulence tends to evolve towards this kind of symmetry.  However, 
on the scale of the largest energy containing eddies, in realistic systems,
we expect that differential shear, background gradients, and 
specific dynamical instabilities may all play a role, and all these effects
will not respect the symmetry between magnetic and velocity fields.
An example of an instability which will necessarily produce such an
asymmetry is the magneto-rotational (or Balbus-Hawley) instability
in accretion disks \cite{v59,c61,bh91}, which is discussed below.

Here we summarize some of the more important conclusions from these arguments.
\begin{enumerate}
\item The fluid helicity is largely irrelevant to flux generation,
except when the field amplitude is very small,
although it does affect flux transport.

\item This prescription eliminates turbulent dissipation for currents
aligned with the large scale magnetic field, but continues to damp
other current components.  This is not a qualitative change in the
role of turbulent damping for most field configurations, but does imply
that force-free large scale fields are protected against turbulent
dissipation.

\item The anomalous magnetic helicity current, $j_H$, depends on
$\langle([{\bf B}\cdot\nabla] {\bf v})({\bf B}\cdot[\nabla\times{\bf v}])\rangle$,
which has no particular relationship to the fluid helicity and is
parity invariant.  Rather than violating spatial symmetry along all three
principal axes, a successful dynamo can result from a situation where
only two out of three directions have broken symmetries. An example of
this is differential rotation in a vertically uniform cylinder.

\item The `$\alpha-\Omega$' dynamo has an analog in this new theory, which
gives a similar growth rate, but which does not depend on any 
background vertical structure.  (The analogous effect is provided by
vertical gradients in $\langle{\bf B}\rangle$.)  In an accretion disk,
the success of the dynamo is tied to the outward transport of 
angular momentum \cite{vc01}.

\item The analog of the `$\alpha^2$' dynamo of conventional theory has the difficulty
that the turbulent dissipation term is of the same order as the driving term
resulting from magnetic helicity transport.  This does not imply that
this kind of dynamo is impossible, but configurations with force-free, or nearly
force-free, fields are strongly favored.

\item The analog of the `$\alpha-\Omega$ dynamo has a strong vertical magnetic
helicity current, which has the same sign as $-\partial_r\Omega(r)$, i.e. 
towards $\hat z$ for an accretion disk and $-\hat z$ for a star like the Sun.
This implies the necessity for magnetic helicity ejection at the system boundaries.
While the energy budget for this is small compared to the energy budget of the
dynamo in an accretion disk, it represents an unsolved aspect of the dynamo
physics.  
\end{enumerate}

We note that the ejection of magnetic helicity from rotating systems as a necessary
part of the dynamo process has been suggested by other authors \cite{bf00a,bf00b}, 
although the terms of the discussion were somewhat different.
In particular, they were concerned with removing the magnetic helicity constraint
by removing the magnetic helicity.  Although realistic systems
can work this way, there is no fundamental reason why the magnetic helicity
can't circulate within a closed system and produce a 
dynamo effect by being carried by large scale fields in one part of the
system and eddy scale fields in another.

\section{Applying and Testing the Theory}

Both reconnection and the dynamo are the subject of intensive experimental research.
Magnetic reconnection is being studied on several dedicated
experiments around the world (MRX, TS3/4, SSX, VTF). In each experiment, 
magnetized
loops are generated and merged. At present, the physical scales of such experiments
are 0.1 to 1 m and the Lundquist number is about 1000. Sophisticated diagnostics
are used to get plasma and magnetic field parameters in these experiments. This
enables testing theoretical predictions. The direct relation between those
experiments and astrophysics is complicated by the fact that some of the
effects that are important in laboratory, e.g. anomalous resistivity, may
not be important in conditions of astrophysical plasma, e.g. interstellar gas.

Dynamo experiments, e.g. using liquid sodium, are mostly focused on the
reproduction of the dynamo effect for the low Lundquist numbers. In this
regime, neither reconnection nor magnetic helicity are expected
to provide strong constraints on the evolution of the experiments.

On the other hand, numerical simulations are a valuable source of information 
for mean-field dynamo theory.  We have already discussed their role in 
undermining the conventional approach to this topic.  It is also important
to note that there are a large number of simulations which seem to show
the operation of a successful dynamo, in the sense that they demonstrate
the growth of a magnetic field with a significant component at large
spatial scales.  These are the simulations of magnetized, 
ionized accretion disks
(e.g. \cite{hb91,hgb95,hgb96,shgb96,bnst95,bnst96} see also \cite{bh98,hb99} for a
review) which are subject to the magneto-rotational instability.  
These simulations include extremely
simplified physics and cover a limited set of spatial scales and
geometries.  Nevertheless, they agree in a number of
important aspects, namely:

\begin{enumerate}
\item Any magnetic seed field undergoes substantial amplification to
a final state which is (apparently) independent of initial conditions.
In this state the magnetic field pressure is a few percent of the
gas pressure and the dimensionless `viscosity', $\alpha_{SS}$ is
about half of this ratio.  
A large fraction of the magnetic energy
is contained in a large scale field with a domain size which is
a large fraction of the simulation box size.  This large scale
field is not static, but varies on time scales of tens of shearing
times, a feature which seen in all of the simulations cited above.

\item
The growth in field strength is rapid, i.e. a significant fraction of
the shear rate, even when the field is weak.  The growth is several
times slower for an initially azimuthal field, where the amplification depends
on dynamo action, as for an initially vertical field, where the amplification
can simply reflect the linear growth of the instability, which generates
azimuthal field, but the point remains true in either case.

\item The magnetic field pressure in the saturated state is not a
constant fraction of the ambient pressure, but varies from a small
fraction at the midplane (for models with vertical structure) to
a value comparable to the gas pressure a few scale heights away from
the midplane.  This effect  is particularly dramatic in the recent
simulations of Miller and Stone \cite{ms00}.
This distribution does not seem to be due to magnetic
buoyancy \cite{shgb96}, that is, the fields are mostly generated locally.
The time averaged magnetic stress ($\langle B_rB_\theta\rangle$) is 
somewhat more uniform.

\item The dynamo persists when vertical gravity is turned off, although
the simulations are slightly different in this case 
\cite{hb92,hgb95}.
\end{enumerate}

A possible interpretation of these results is that the simulations
are showing a chaotic dynamo \cite{b50,k67}, in which turbulent stretching of embedded 
magnetic field lines results in a runaway amplification of the magnetic field.
One problem with this is that the accretion disk simulations are unique in 
generating substantial magnetic field energy on scales much larger than the
typical eddy size.  Simulations of MHD turbulence in a box typically
produce magnetic field structure whose energy spectrum peaks on scales slightly 
{\it smaller\ } than a typical eddy scale and with a total energy density 
which is a (large) fraction of the kinetic energy density.  At longer 
wavelengths the magnetic energy density falls, although slowly.
Another point is that the vertical distribution of magnetic energy is not
simply a reflection of local conditions, but seems to show some sort of
global field evolution.  The obvious conclusion 
is that some sort of large scale dynamo effect is being produced
in the simulations, as a consequence of the Balbus-Hawley instability.
Since there has been no attempt to
look specifically at the flow of magnetic helicity, it is difficult to know 
whether to ascribe the dynamo effect to a locally produced fluid helicity 
(in which case the dynamo should slow down as the resolution is increased) or 
to a turbulently driven magnetic helicity current.  The dynamo growth rate does 
not {\it appear} to slow down in the higher resolution studies, but this has not 
been examined critically. Further study of these simulations should 
allow testing of the notion that the turbulently driven magnetic helicity
current is playing a critical role in these simulations.

There is one indirect test which has already been performed.  Equation (\ref{jh2})
can be used to show, via integration by parts, that the sign of the
magnetic helicity current depends on the direction of angular momentum
transport.  Reversing the sign of the magnetic helicity current has the effect
of turning off the dynamo.  One simple numerical experiment is to conduct
a simulation in which the angular momentum current flows in the opposite
direction.  This has been done \cite{hgb96} by turning off the centrifugal
force term, so that the turbulence is driven only by a kind of magnetized
Kelvin-Helmholtz instability.  The dynamo effect was suppressed and the magnetic
field decayed away, after an initial burst of growth.  
This simulation had a limited dynamic range,
so that all the eddies were dominated by the local shear. Consequently,
the elimination of the dynamo effect led to a complete suppression of
the magnetic field through azimuthal stretching and radial mixing.

There
has been a recent attempt to combine shear with an asymmetrically driven
turbulence \cite{bbs01} to produce a non-heliacal dynamo.  The results
were disappointing.  The expected correlation between the magnetic
helicity flux and the velocity correlation seen in equation (\ref{jh2})
was found, but the magnetic helicity flux was largely divergenceless
and there was no clear correlation between the its small divergence and
the electromotive force.  This may have been due to the boundary
conditions, which forced a return loop of magnetic flux within the box.
Clearly further numerical experiments would be useful.

Assuming that we can understand the conceptual basis of accretion disk
dynamos, it should be possible to construct a useful mean-field theory
that incorporates transport effects and allows us to predict the
dynamics of accretion disk fields.  This model will need to incorporate
the effects of fluctuations in the electromotive force \cite{vb97}.  In
the conventional mean-field dynamo theory such fluctuations have been
shown to be capable of driving a mean-field dynamo in the absence of
any average helicity.  Their role in the modified version of mean-field
dynamo theory is not yet understood.  Such a model would be useful for
building models of disks that incorporate both realistic local physics
and MHD turbulence.  A similar effort should be made for stellar dynamos,
although there has been, as yet, no progress in this direction.  There
has been work on the galactic dynamo \cite{k02} which incorporates the
notion of magnetic helicity current, including the term given in
equation (\ref{jh1}).

Much less progress has been made in numerical simulations of stochastic
reconnection.  This is particularly unfortunate since
magnetic reconnection is one of the most fundamental properties of the magnetic
field dynamics in the conducting fluid, and its applications are not limited
to its consequences for astrophysical dynamos.  In fact,
reconnection is likely to be extremely important for the dynamics of the advection
dominated flows, star formation, propagation (see \cite{cly02})
and acceleration (see \cite{dl00,dl01}) of cosmic rays, dynamics of
charged dust \cite{ly02}.
Direct study of the reconnection layer is difficult as both very small
scales (turbulent microscales comparable to the current sheet thickness) 
and large scales (the contact region scale)
are present in the problem. The requirement that we evolve structures at all 
scales over the whole broad reconnection region suppress any hope that adaptive
mesh codes would be very helpful.
Nevertheless, simple diagnostics may be used to
distinguish fast stochastic reconnection from the Sweet-Parker
model.  
For instance, using MHD simulations we can measure currents $J$, magnetic fields
$B$ and velocities $v$.
If we divide $\langle J^2[(B\cdot\nabla)B\cdot v]^2\rangle$
 by $\langle J^2(\nabla \times B)^2\rangle$ then
we have a measure of the rms magnetic field across a typical current sheet
times the speed with which it is expelled. An approximate measure of
reconnection speed can then be obtained by dividing the result by 
$\langle (J\times B)^2 \rangle/\langle J^2\rangle$ and taking the square root.
Although reconnection rates for low Lundquist numbers are not so different
for the Sweet-Parker and stochastic reconnection models, the scaling of the
reconnection rates with the Alfven velocity are very different. This gives
some
hope that the stochastic reconnection model can be tested before long.

\section{Discussion and Summary}

It is not possible to understand the astrophysical dynamo
and dynamics of magnetized astrophysical 
plasmas without understanding how magnetic fields evade the
topological constraints imposed by flux-freezing.  This obviously
includes the problem of reconnection, but also the more subtle
difficulty posed by magnetic helicity conservation.  Here we
have compared traditional approaches to the problem
of magnetic reconnection and the mean-field dynamo with new approaches based
on an explicit recognition of the role geometry plays in both these
problems.  In fact,  one of the more striking aspects of stochastic
reconnection model \cite{lv99}is that the global
reconnection speed is 
relatively insensitive to the actual physics of reconnection.
Equation (\ref{main}) only
depends on the nature of the turbulent cascade. 
In the end, reconnection
can be fast because if we consider any particular flux element
inside the contact volume, assumed to be of order $L_x^3$,
the fraction of the flux element that actually undergoes
microscopic reconnection vanishes as the resistivity goes
to zero.  This is turn implies that reconnection is not tightly
coupled to electron heating.
More generally, the results presented here suggest that, in most
cases, microphysics is irrelevant to the dynamo process.

Although objections to conventional dynamo theory tend to conflate
the issues of reconnection and magnetic helicity conservation,
these are, in fact, two separate problems, for which we have
proposed two separate resolutions.  Taken together, they imply
that astrophysical dynamos are capable of operating in a broad
range of circumstances.  However, it is important to remember that
they stand separately.  Conventional dynamo theory is not rescued
by assuming rapid reconnection, although it requires it.  Conversely,
the use of equations (\ref{div}) and (\ref{jh1}) to describe
dynamo activity do not require stochastic reconnection, but only that
{\it some} model of fast reconnection work.

Our main conclusions are as follows:

\begin{itemize}
\item The rate of magnetic reconnection is increased dramatically
in the presence of a stochastic component to the magnetic field.
Even when the turbulent cascade is weak the resulting reconnection
speed is independent of the Ohmic resistivity.  However, it is extremely
sensitive to the level of noise.  This may explain the variable rates
of magnetic reconnection seen in the solar corona.  It also implies
that laminar flow patterns that drive a magnetic helicity current may
still require some level of local turbulence in order to drive a 
large scale dynamo.

\item The argument that the rapid rise of random magnetic field associated
with dynamo action results in the  suppression of dynamo \cite{ka92}
is untenable since the increase of the random component
of the magnetic field increases the reconnection rate. We conclude that
dynamo is a self-regulating process.

\item Conventional mean-field dynamo theory, which does not account for the 
conservation of
magnetic helicity is ill-founded. The suggested modification of the mean-field
dynamo equations allow us to account for results of numerical sumulations
and make the theory, for the first time ever, self-consistent.

\end{itemize}

{\bf Acknowledgements}. AL and JC acknowledge the support of NSF grant
NSF AST-0125544. 
ETV acknowledges the support of NSF grant AST-0098615.
AL thanks the LOC for the financial support.

\end{document}